\documentclass[oneside,reqno,a4paper,11pt]{amsart}

\usepackage{color,soul,supertabular,longtable,verbatim}
\usepackage{amsmath,bbm,amsfonts,amssymb}
\usepackage{hyperref}
\usepackage{bibtopic}
\newtheorem {theorem}{Theorem}[section]
\newtheorem {lemma}[theorem]{Lemma}

\newtheorem {proposition}[theorem]{Proposition}
\theoremstyle{remark}

\theoremstyle{problem}

\theoremstyle{definition}

\theoremstyle{plain} \numberwithin {equation}{section}

\def\XXint #1#2#3{{\setbox 0=\hbox {$#1{#2#3}{\int }$}
\vcenter {\hbox {$#2#3$}}\kern -.5\wd 0}}

\def\ba{\begin{aligned}}
\def\ea{\end{aligned}}
\def\bn{\begin{enumerate}}
\def\en{\end{enumerate}}
\def\be{\begin{equation}}
\def\ee{\end{equation}}

\def\lap{\triangle}

\def\g{\gamma}

\def\O{\Omega}

\def\ep{\epsilon}

\def\R{\mathcal{R}}

\def\p{\partial}

\def\norm[#1]#2{\|#2\|_{#1}}
\def\p{\partial}

\begin{document}




\vspace{1cm}

\title{A Blow-Up Criterion for classical solutions to the Compressible Navier-Stokes equations}
\author[Xiangdi Huang, Zhouping Xin]
{Xiangdi Huang, Zhouping Xin}

\address{ Xiangdi Huang \hfill\break\indent
          The Institute of Mathematical Sciences, The Chinese University of Hong Kong, Hong Kong}
          \email{\href{mailto:xdhuang@math.cuhk.edu.hk}{xdhuang@math.cuhk.edu.hk}}
\address{ Zhouping Xin \hfill\break\indent
          The Institute of Mathematical Sciences, The Chinese University of Hong Kong, Hong Kong}
          \email{\href{mailto:zpxin@ims.cuhk.edu.hk}{zpxin@ims.cuhk.edu.hk}}
\maketitle

\begin{abstract}
In this paper, we obtain a blow up criterion for classical solutions to
the 3-D compressible Naiver-Stokes equations just in terms of the gradient
of the velocity, similar to the Beal-Kato-Majda criterion
for the ideal incompressible flow.
  In addition, initial vacuum is allowed in our case.
\end{abstract}


\section{introduction}

Let $\O\subset \R^n$ be a $n$-dimensional domain. The time
evolution of the density and the velocity of a general viscous
compressible barotropic fluid occupying a domain $\O$ is governed by
the Navier-Stokes system of equations
\begin{equation}
\left\{ \ba
& \p_t\rho + \rm{div}(\rho u)=0,\\
& \p_t(\rho u) + \rm{div}(\rho u\otimes u) -\mu\lap u-(\mu +
\lambda)\nabla(\rm{div}u) + \nabla P(\rho)=0 \ea \right.
\end{equation}

Where $\rho, u, P$ denotes the density, velocity and pressure respectively.
The equation of state is given by
\be
 P(\rho) = a\rho^{\g}\quad (a>0,\g>1)
 \ee
$\mu$ and $\lambda$ are the shear viscosity and the bulk viscosity coefficients
respectively satisfying the condition:
\be
\mu>0, \lambda + \frac{2}{3}\mu\ge 0
\ee

   Lions \cite{L1} \cite{L2}, Feireisl \cite{F1}\cite{F2} et. established the global existence of
   weak solutions to the problem $(1.1)-(1.3)$, where vacuum is allowed initially.
    The existence of global smooth solutions to the compressible Navier-Stokes equations is
   obtained by Matsumura\cite{M1}and Nishida under the condition that the initial data is a small perturbation
of a non-vacuum constant. It is also shown by Xin\cite{X1} that there is no global
in time regular solution in $R^3$ to the compressible Naiver-Stokes equations provided that the
initial density is compactly supported.

There are many results concerning the existence of strong solutions
to the Navier-Stokes equations, only local existence results have been established, see \cite{K1},
   \cite{K2},\cite{K3},\cite{S2}. V.A.Solonnikov proved in \cite{S1}
that for $C^2$ pressure laws and initial data satisfies for some
$q>N$, \be 0<m\le\rho_0(x)\le M<\infty,\quad and\quad \rho_0\in
W^{1,q}(T^N) \ee

\be u_0\in W^{2-\frac{2}{q},q}(T^N)^N \ee there exists a local unique
strong solution $(\rho,u)$ to $(1.4)-(1.5)$ for periodic data, such that
\be
\ba
& \rho\in L^{\infty}(0,T;W^{1,q}(T^N)),\quad \rho_t\in L^q((0,T)\times T^N)\\
& u\in L^q(0,T;W^{2,q}(T^N)),\quad u_t\in L^q((0,T)\times T^N)^N
\ea
\ee

Later, it was shown in \cite{K1} that if $\O$ is either a bounded
domain or the whole space, the initial data $\rho_0$
and $u_0$ satisfy \be 0\le\rho_0\in W^{1,\tilde{q}}(\O),\quad u_0\in
H_0^1(\O)\cap H^2(\O) \ee for some $\tilde{q}\in (3,\infty)$ and the
compatibility condition:
\be -\mu\triangle u_0 - (\lambda +
\mu)\nabla\rm{div}u_0 + \nabla P(\rho_0) = \rho_0^{1/2}g \quad
for\quad some\quad g\in L^2(\O)
\ee
then there exists a positive time $T_1\in (0,\infty)$ and a unique
strong solution $(\rho,u)$ to the isentropic problem, such that
\be
\ba
& \rho\in C([0,T_1];W^{1,q_0}(\O))\\
& u\in C([0,T_1];D_0^1\cap D^2(\O))\cap L^2(0,T_1;D^{2,q_0}(\O))
\ea
\ee

 Furthermore, one has the
following blow-up criterion: if $T^*$ is the maximal time of
existence of the strong solution $(\rho,u)$ and $T^*<\infty$, then
\be \sup\limits_{t\rightarrow T^*}(\norm[W^{1,q_0}]{\rho} +
\norm[D_0^1]{u}) = \infty \ee where $q_0 = \min (6,\tilde{q})$.

Here and throughout this paper, we use the following notations for the standard
 homogeneous and inhomogeneous Sobolev spaces.

$$ D^{k,r}(\O) =\{u\in L^1_{loc}(\O):\norm[L^r]{\nabla^k u}<\infty\}, $$
$$W^{k,r} = L^r\cap D^{k,r}, H^k = W^{k,2},\quad D^k = D^{k,2}$$
$$D_0^1 = \{u\in L^6(\O):\norm[L^2]{\nabla u}<\infty\quad and\quad u=0\quad on\quad \p\O\},$$
$$ H_0^1 = L^2\cap D_0^1,\quad \norm[D^{k,r}]{u} = \norm[L^r]{\nabla^k u}$$

 Recently, it is established in \cite{J1} that
 if $[0,T^*)$ is the finite maximal interval for such strong solutions. and $7\mu>9\lambda$, then
\be
 \lim_{T\rightarrow T^*}\{\sup\limits_{0\le
T< T^*}\norm[L^{\infty}]{\rho} + \int_0^T(\norm[W^{1,q_0}]{\rho} +
\norm[L^2]{\nabla\rho}^4)dt\} = \infty
\ee
Here they only require a sufficient regularity of density $\rho$ to
 admit the global existence of strong solutions, as $(1.11)$ revealed.

  It is shown in \cite{H2}, we can obtain a blow up criterion for strong solutions similar to
  Beal-Kato-Majda for ideal incompressible fluid, i.e,
$$
\int_0^{T_*}\norm[L^{\infty}]{\nabla u}dt = \infty
$$
  where we assume that
  \be
\ba
& \mu + \lambda\ge 0, \quad N=2, \quad \O = T^2 \\
& \mu + \lambda = 0, \quad N=3,\quad \O\subset R^3
\ea
\ee

  Recently, it is shown in\cite{K3} that if the domain is either a bounded domain
  or the whole space $R^3$ and the initial data $\rho_0,u_0$ satisfy
\be
\ba
& (\rho_0,P_0)\in H^3, \rho_0\ge 0 \\
& u_0\in H_0^1\cap H^3
\ea
\ee
and the compatibility condition
\be
-Lu_0 + \nabla P(\rho_0) = \rho_0g \quad
for\quad some\quad g\in H_0^1(\O)\quad with\quad \rho_0^{\frac{1}{2}}g\in L^2
\ee
where
$$
Lu = \mu\triangle u + (\lambda +
\mu)\nabla\rm{div}u, \quad P(\rho_0) = a\rho_0^{\g}
$$
Then there exist a small time $T_*\in (0,T)$ and a unique classical solution $(\rho,p,u)$
such that
\be
\ba
& (\rho,P)\in C([0,T_*];H^3(\O))\\
& u\in C([0,T_*];D_0^1\cap D^3(\O))\cap L^2(0,T_*;D^4(\O))\\
& u_t\in L^{\infty}(0,T_*;D_0^1(\O))\cap L^2(0,T_*;D^2(\O)) \quad and\quad
 \sqrt{\rho}u_t\in L^{\infty}(0,T_*;L^2(\O)) \\
& (u_t,\nabla^2u)\in C((0,T_*]\times\bar{\O})
\ea
\ee

  In this paper, under the assumption
\be
\mu>\frac{1}{7}\lambda
\ee
  we establish a blow up criterion for classical solutions.

Here and thereafter $C$ always denotes a generic constant depending only on
$\O,T$ and initial data.

For the initial boundary value problem, we have the following
result:

\begin{theorem}
Let $\O\subset R^3$ be a bounded domain. $Q_T = (0,T)\times\O$. Assume that the initial data
satisfy $(1.13)-(1.14)$. Let $(\rho,u)$ be a classical solution of
the problem $(1.1)-(1.3)$ satisfying the regularity $(1.15)$. If $T^*<\infty$ is the
maximal time of existence, then \be \lim_{T\rightarrow
T^*}\int_0^T\norm[L^{\infty}(\O)]{\nabla u}dt = \infty \ee
provided that $(1.16)$ holds.
\end{theorem}

In case of the Cauchy problem, it holds that
\begin{theorem}
  Let $\O = R^3$. Assume that the initial data satisfy
\be
(\rho_0,P_0)\in H^3(R^3),\quad u_0\in
D_0^1(R^3)\cap D^3(R^3)
\ee
the compatibility condition $(1.14)$. Let $(\rho,u)$ be a classical solutions to the
problem $(1.1)-(1.3)$ in the sense of \cite{K3} satisfying
\be
\ba
& (\rho,P)\in C([0,T_*],H^3(R^3)) \\
& u\in C([0,T_*],D_0^1(R^3)\cap D^3(R^3))\cap L^2(0,T_*;D^4(R^3))\\
& u_t\in L^{\infty}(0,T_*;D_0^1(R^3))\cap L^2(0,T_*;D^2(R^3)),\quad \sqrt{\rho}u_t\in
L^{\infty}(0,T_*;L^2(R^3))
\ea
\ee
If $T^*<\infty$ is the maximal time of existence, then \be
\lim_{T\rightarrow T^*}\int_0^T\norm[L^{\infty}(\O)]{\nabla u}dt =
\infty \ee
provided that $(1.16)$ holds.
\end{theorem}

{\bf Remark 1.1} The blow up criterion $(1.10)$ involves both the density and velocity. It may be
natural to expect the higher regularity of velocity if the density is regular enough. $(1.11)$ shows
that sufficient regularity of the gradient of density indeed guarantees the global existence of strong
solutions. The main difficulty in our case is to control the gradient of density, which is not a
priorily known and coupled with the second derivative of velocity.

  In this paper, we establish a blow up criterion under condition $(1.16)$ instead of $(1.12)$.
  Obviously, $(1.16)$ becomes physical condition $(1.3)$ if $\lambda\le0$.
     We develop some new estimates under the condition that the integral on the left
     of $(1.17)$ is finite. In fact, the
key estimate in our analysis is $L^{\infty}H^1$ bound of $\nabla\rho$.
 To control the $L^{\infty}(0,T;L^2(\O))$ norm of $\nabla\rho$, we observe that
 that the space-time square mean of the convection term $F = \rho u_t + \rho u\cdot\nabla u$
 is controlled by that of $\nabla\rho$ (see Lemma 2.3). This, in turn, gives the desired
  $L^{\infty}(0,T;L^2(\O))$ estimate on $\nabla\rho$, and thus the $L^2(0,T;H^2(\O))$ of $u$.
  To obtain a higher regularity of $\nabla\rho$, one need to improve the regularity of pressure $P$,
  as we can't deduce $P\in L^{\infty}H^3$ directly even
  $\rho$ is sufficient regular unless $\g=2$ or $\g\ge 3$ due to the presence of vacuum.
  Our proof relies on the observation that, the pressure $P$ is solution of a transport equation
  $P_t + \rm{div}(Pu) + (\g-1)P\rm{div}u = 0$. Hence we can deduce a high regularity of $P$ provided
  that $u$ and $P_0$ are regular enough. As a consequence, the high order regularity of the density follows
  from the mass equation and a sufficient regularity of pressure.

{\bf Remark 1.2} There are many results concerning blow-up criteria of the incompressible flows.
In the well-known paper \cite{B1}, Beal-Kato-Majda established a blow-up
criterion for the incompressible Euler equations. One can get global smooth solution if
 $\int_0^T\norm[L^{\infty}]{\omega}dt$ is bounded. It's worth noting
 that only the vorticity $\omega$ plays an important role in the
  existence of global smooth solutions. Moreover, as pointed out by Constantin\cite{C5},
 the solution is smooth if and only if $\int_0^T\norm[L^{\infty}]{((\nabla u)\xi)\cdot \xi}$ is bounded,
 where $\xi$ is the unit vector in the direction of $\omega$. It turns out that the solution
 becomes smooth either the asymmetric
 or symmetric part of $\nabla u$ is controlled.
   Later, Constantin\cite{C3}, Fefferman and Majda
 showed a sufficient geometric condition to control the breakdown of smooth solutions of incompressible
 Euler involving the Lipschitz regularity of the direction of the vorticity. It is also shown by
 Constantin\cite{C4} and Fefferman that the solution of incompressible Navier-Stokes equations is smooth if
 the direction of vorticity is well behaved.

  Recently, in\cite{C1}, assuming that the added stress tensor is given in a proper form,
   and using an idea of J.-Y. Chemin and N. Masmoudi \cite{C2},
 Constantin, P. and Fefferman, C., Titi, E. S. and Zarnescu, A
 obtain a logarithmic bound for $\int_{0}^{T}\|\nabla u\|_{L^{\infty}}dt.$ to conclude
 that the solution to Navier-Stokes-Fokker-Planck system exists for all time and is smooth.

 In our paper, we establish a similar criterion to Beal-Kato-Majda. Our blow up criteria involve
 both the symmetric and asymmetric part of $\nabla u$, as the compressibility and the vorticity
 of the compressible flow are
 two key issues in the formation of singularities of the compressible Navier-Stokes.

{\bf Remark 1.3} The paper is organized as follows. Section 2 is devoted to improve the
regularity of the density and the velocity in strong sense. In section 3, we derive some high
order regularity estimate for the density, pressure and velocity,
which guarantee the extension of classical solutions.

\section{Regularity of the density and the velocity}

Let $(\rho,u)$ be a classical solution to the problem $(1.1)-(1.3)$. We
assume that the opposite  holds, i.e
\be \lim_{T\rightarrow
T^*}\int_0^T\norm[L^{\infty}(\O)]{\nabla u}dt\le C<\infty \ee
First, the standard energy estimate yields \be \sup\limits_{0\le t\le
T}\norm[L^2]{\rho^{1/2}u(t)} + \int_0^T\norm[H^1]{u}^2dt \le C,\quad 0\le
T<T^* \ee

By assumption $(2.1)$ and the conservation of mass, the $L^{\infty}$ bounds of density
follows immediately,
\begin{lemma}
  Assume that
\be
  \int_0^T\norm[L^{\infty}]{div u}dt\le C, \quad 0\le T<T^*
\ee

then  \be
  \norm[L^{\infty}(Q_T)]{\rho} \le C,\quad 0\le T<T^*
\ee
\end{lemma}
\proof It follows from the conservation of mass that for $\forall q>1$,
\be \p_t(\rho^q) +
\rm{div}(\rho^qu) + (q-1)\rho^q\rm{div}u = 0 \ee
Integrating $(2.5)$ over $\O$ to obtain,
\be \p_t\int_{\O}\rho^qdx  \le
(q-1)\norm[L^{\infty}(\O)]{\nabla u}\int_{\O}\rho^qdx \ee i.e \be
\p_t\norm[L^q]{\rho} \le \frac{q-1}{q}\norm[L^{\infty}(\O)]{\nabla
u}\norm[L^q]{\rho} \ee
 which implies immediately
  \be
\norm[L^q]{\rho}(t)\le C \ee
with $C$ independent of $q$, so our lemma follows.

\endproof

Next, we improve the energy estimate $(2.2)$. It's worth noting that only here we
require that the condition $(1.16)$ holds.
\begin{lemma}
Let $\mu>\frac{1}{7}\lambda$, then there exist a small $\delta>0$
\be
\sup_{0\le t\le T}\int_{\O}\rho|u|^{3+\delta}dx\le C,\quad 0< T< T_*,
\ee
where $C$ is a positive constant depending only on $\norm[L^{\infty}(Q_T)]{\rho}$.
\end{lemma}

\proof
This follows from an argument due to Hoff\cite{H3}.

Indeed, setting $q>3$ and multiplying $(1.2)$ by $q|u|^{q-2}u$, and integrating over $\O$, we obtain
by using lemma $2.1$ that
\be
\ba
& \frac{d}{dt}\int_{\O}\rho|u|^qdx + \int_{\O}(q|u|^{q-2}[\mu|\nabla u|^2 +
(\lambda + \mu)(\rm{div}u)^2 + \mu(q-2)|\nabla|u||^2] \\
& + q(\lambda + \mu)(\nabla|u|^{q-2})\cdot u\rm{div}u)dx \\
& = q\int_{\O}\rm{div}(|u|^{q-2}u)pdx\\
& \le C\int_{\O}\rho^{\frac{1}{2}}|u|^{q-2}|\nabla u|dx\\
& \le\ep\int_{\O}|u|^{q-2}|\nabla u|^2dx + C(\ep)\int_{\O}\rho|u|^{q-2}dx\\
& \le\ep\int_{\O}|u|^{q-2}|\nabla u|^2dx + C(\ep)(\int_{\O}\rho|u|^qdx)^{\frac{q-2}{q}}
\ea
\ee

Note that $|\nabla |u||\le |\nabla u|$, one gets that
\be
\ba
& q|u|^{q-2}[\mu|\nabla u|^2 + (\lambda + \mu)(\rm{div}u)^2 + \mu(q-2)|\nabla|u||^2]
 + q(\lambda + \mu)(\nabla|u|^{q-2})\cdot u\rm{div}u \\
 & \ge q|u|^{q-2}[\mu|\nabla u|^2 + (\lambda + \mu)(\rm{div}u)^2 + \mu(q-2)|\nabla|u||^2\\
 & - (\lambda + \mu)(q-2)|\nabla|u||\cdot|\rm{div}u|]\\
 & = q|u|^{q-2}[\mu|\nabla u|^2 + (\lambda + \mu)(\rm{div}u - \frac{1}{2}|\nabla|u||)^2]\\
 & + q|u|^{q-2}[\mu(q-2) - \frac{1}{4}(\lambda + \mu)(q-2)^2]|\nabla|u||^2\\
 & \ge C|u|^{q-2}|\nabla u|^2
\ea
\ee
where we use the fact $\mu>\frac{1}{7}\lambda$, then there exist a small $\delta$, such that
\be
\mu(q-2) - \frac{1}{4}(\lambda + \mu)(q-2)^2>0
\ee
where $q=3+\delta$.

Inserting $(2.11)$ into $(2.10)$, and taking $\ep$ small enough, we may apply
Gronwall's inequality to conclude $(2.9)$.

\endproof

The next lemma shows a connection between a convection term and
the gradient of the density, which will play an important role in deriving the desired
bounds on $\nabla\rho$.
\begin{lemma}
  Let $F=\rho u_t + \rho u\cdot\nabla u$. Then it holds that
  $$\int_{Q_T}F^2dxdt + \sup_{0\le t\le T}\int_{\O}|\nabla u|^2dx
   \le C\int_{Q_T}|\nabla\rho|^2dxdt + C,\quad 0\le T<T^*$$
\end{lemma}

\proof

Note that
\be
\int_{Q_T}F^2dxdt \le C^*(\norm[L^{\infty}(Q_T)]{\rho})\int_{Q_T}\rho
u_t^2dxdt +
 2\int_{Q_T}|\rho u\cdot\nabla u|^2dxdt
\ee

The last term of $(2.13)$ can be estimated as follows
\be
\ba
& \int_{Q_T}\rho^2 |u|^2|\nabla u|^2dxdt \\
&\le C\int_0^T\norm[L^4]{\rho^{\frac{1}{2}}u}^2\norm[L^4]{\nabla u}^2dt\\
&\le C\int_0^T\norm[L^{3+\delta}]{\rho^{\frac{1}{2}}u}^{\alpha}\norm[L^6]{\rho^{\frac{1}{2}}u}^{2-\alpha}
\norm[L^4]{\nabla u}^2dt\\
&\le C\int_0^T\norm[L^2]{\nabla u}^{3-\alpha}\norm[L^{\infty}]{\nabla u}dt\\
&\le C\sup_{0\le T<T^*}\norm[L^2]{\nabla u}^{3-\alpha}\int_0^T\norm[L^{\infty}]{\nabla u}dt\\
&\le C\sup_{0\le T<T^*}\norm[L^2]{\nabla u}^{3-\alpha}
\ea
\ee
where $\frac{\alpha}{3+\delta} + \frac{2-\alpha}{6} = \frac{1}{2}$ and $1<\alpha = \frac{3+\delta}{3-\delta}<2$.
It follows from $(2.13)$ and $(2.14)$ that
\be
\ba
\int_{Q_T}F^2dxdt & \le C^*(\norm[L^{\infty}(Q_T)]{\rho})\int_{Q_T}\rho
u_t^2dxdt + C\sup_{0\le T<T^*}\norm[L^2]{\nabla u}^{3-\alpha}
\ea
\ee

 Multiplying the momentum equation by $u_t$ and integrating show that
\be
\int_{\O}\rho u_t^2dx + \int_{\O}\rho u\cdot\nabla u\cdot u_tdx +
\frac{d}{dt}\int_{\O}\frac{\mu}{2}|\nabla u|^2 +
\frac{\lambda + \mu}{2}(\rm{div}u)^2dx = \int_{\O}P\rm{div}u_tdx
\ee

Note that
\be
\int_{\O}P\rm{div}u_tdx = \frac{d}{dt}\int_{\O}P\rm{div}udx -
\int_{\O}P_t\rm{div}udx,
\ee
and
\be
P_t + \rm{div}(Pu) + (\g-1)P\rm{div}u = 0
\ee
One gets
\be
\ba \int_{\O}P\rm{div}u_tdx & = \frac{d}{dt}\int_{\O}P\rm{div}udx +
\int_{\O}\rm{div}(Pu)\rm{div}udx +
(\g-1)\int_{\O}P(\rm{div}u)^2dx\\
& = \frac{d}{dt}\int_{\O}P\rm{div}udx - \int_{\O}(Pu)\cdot\nabla\rm{div}udx
+ (\g-1)\int_{\O}P(\rm{div}u)^2dx \ea
\ee

This, together with $(2.16)$, yields
\be
\ba & \quad \int_{\O}\frac{\mu}{2}|\nabla u|^2 +
\frac{\lambda + \mu}{2}(\rm{div}u)^2dx(T) + \int_{Q_T}\rho
u_t^2dxdt
 + \int_{Q_T}\rho u\cdot\nabla u\cdot u_tdxdt  \\
& = \int_{\O}\frac{\mu}{2}|\nabla u_0|^2 +
\frac{\lambda + \mu}{2}(\rm{div}u_0)^2dx(T) + \int_{\O}P\rm{div}udx(T)
- \int_{\O}P_0\rm{div}u_0dx\\
& - \int_{Q_T}Pu\cdot\nabla\rm{div}udxdt +
(\g-1)\int_{Q_T}P(\rm{div}u)^2dxdt \ea
\ee

Direct estimates show that
\be
\int_{\O}P\rm{div}udx(T)\le \frac{\mu}{4}\int_{\O}|\nabla u|^2dx(T) +
C
\ee

\be
\ba \int_{Q_T}\rho u\cdot\nabla u\cdot u_tdxdt & \le
\frac{1}{2}\int_{Q_T}\rho u_t^2 +
\int_{Q_T}\rho |u\cdot\nabla u|^2dxdt\\
&\le \frac{1}{2}\int_{Q_T}\rho u_t^2 + C\sup_{0\le T<T^*}\norm[L^2]{\nabla u}^{3-\alpha}
\ea
\ee

On the other hand, it follows from $Lu = F+ \nabla P$ and standard elliptic regularity that
\be
\norm[H^2]{u}\le C(\norm[L^2]{F} + \norm[L^2]{\nabla P})
\ee

\be
\ba
\int_{Q_T}Pu\cdot\nabla\rm{div}udxdt & \le C\norm[L^2]{\rho^{\frac{1}{2}}u}\norm[L^2]{\nabla\rm{div}u}\\
& \le C\int_{Q_T}|\nabla\rho|^2dxdt + \ep\int_{Q_T}F^2dxdt + C
\ea
\ee

Consequently,
\be
\ba
\int_{Q_T}\rho u_t^2dxdt + \frac{\mu}{2}\int_{\O}|\nabla u|^2dx(T) & \le
C\int_{Q_T}|\nabla\rho|^2dxdt + 2\ep\int_{Q_T}F^2dxdt \\
& + C\sup_{0\le T<T^*}\norm[L^2]{\nabla u}^{3-\alpha}
\ea
\ee

Choosing $\ep$ as $2C^*\ep <1$, one concludes that
$$\sup_{0\le t\le T}\int_{\O}|\nabla u|^2dx + \int_{Q_T}F^2dxdt \le C\int_{Q_T}|\nabla\rho|^2dxdt + C$$
This completes the proof of Lemma $2.2$.

\endproof
We are now ready to obtain the desired $L^{\infty}(0,T;L^2(\O))$ estimate of $\nabla\rho$.
\begin{proposition}
 Under the assumption $(2.1)$, it holds that
  \be
  \sup\limits_{0\le t\le T}\int_{\O}|\nabla\rho|^2dx \le C \quad, 0\le T<T^*
  \ee
\be
\int_{Q_T}\rho u_t^2dxdt + \sup\limits_{0\le t\le T}\int_{\O}|\nabla
u|^2dx\le C \quad, 0\le T<T^*
\ee
\be
\int_0^T\norm[H^2(\O)]{u}^2dt\le C,\quad 0\le T<T^*
\ee
\end{proposition}

\proof Differentiating the mass equation in $(1.1)$ with respect to $x_i$,
and multiplying the resulting identity by $2\p_i\rho$ yield
\be
\p_t|\p_i\rho|^2 + \rm{div}(|\p_i\rho|^2u) + |\p_i\rho|^2\rm{div}u +
2\p_i\rho\rho\p_i\rm{div}u + 2\p_i\rho\p_i u\cdot\nabla\rho = 0
\ee
Integrating $(2.29)$ over $\O$ shows that

\be
\ba \p_t\int_{\O}|\p_i\rho|^2dx & =
-\int_{\O}|\p_i\rho|^2\rm{div}udx - 2\int_{\O}\rho\p_i\rho \p_i\rm{div}udx
 - \int_{\O}2\p_i\rho\p_iu\cdot\nabla\rho dx\\
& = -(A_1 + A_2 + A_3) \ea
\ee

Each term on the right hand side of $(2.30)$ can be estimated as follows:

\be
|A_1(t)|\le \norm[L^{\infty}]{div u}(t)\int_{\O}|\p_i\rho|^2dx
\le\norm[L^{\infty}]{div u}(t)\int_{\O}|\nabla\rho|^2dx
\ee
It follows from $(2.22)$ that,
\be
|A_2(t)|\le C\norm[L^2]{\nabla\rho}(\norm[L^2]{\nabla P} + \norm[L^2]{F}) \le
C(\int_{\O}|\nabla\rho|^2dx + \int_{\O}F^2dx)
\ee

\be
|A_3(t)|\le C\norm[L^{\infty}]{\nabla u}(t)\int_{\O}|\nabla\rho|^2dx
\ee

Consequently,
\be
\p_t\int_{\O}|\nabla\rho|^2dx \le C(\norm[L^{\infty}]{\nabla u}(t) +
1) \int_{\O}|\nabla\rho|^2dx + C\int_{\O}F^2dx
\ee

This, together with Gronwall's inequality, yields
\be
\ba
& \int_{\O}|\nabla\rho|^2dx(t) \\
& \le Ce^{C\int_0^t(\norm[L^{\infty}]{\nabla u}(s) + 1)ds}
(\int_{\O}|\nabla\rho_0|^2dx + \int_0^t(\int_{\O}F^2(s)dx)
e^{-C\int_0^s(\norm[L^{\infty}]{\nabla u}(\tau) + 1)d\tau}ds)\\
&\le C\int_0^t\int_{\O}F^2dxds + C\\
&\le C\int_0^t\int_{\O}|\nabla\rho|^2dxds + C \ea
\ee

Hence
\be
\sup\limits_{0\le t\le T}\int_{\O}|\nabla\rho|^2dx \le C
\ee

Next, it follows from $(2.15),(2.24)$ and $(2.35)$ that
\be
\int_{Q_T}\rho u_t^2dxdt + \sup\limits_{0\le t\le T}\int_{\O}|\nabla
u|^2dx\le C
\ee
This, together with $L u = \rho u_t + \rho u\cdot\nabla u +
\nabla P$, shows that

\be
\ba \norm[L^2(0,T;H^2(\O))]{u} & \le \norm[L^2(Q_T)]{\rho u_t} +
\norm[L^2(Q_T)]{\rho u\cdot\nabla u} +
\norm[L^2(Q_T)]{\nabla P} \\
& \le C + C\norm[L^2(Q_T)]{\nabla\rho}\le C \ea
\ee

\endproof

Next, we proceed to improve the regularity of $\rho$ and
$u$. To this end, we first derive some bounds on derivatives of $u$ based on
above estimates.

\begin{proposition}
 Under the condition $(2.1)$, it holds that
  \be
  \sup\limits_{0\le t\le T}\norm[L^2]{\rho^{1/2}u_t(t)}^2 + \int_{Q_T}|\nabla u_t|^2dxdt \le C,
  \quad 0\le T<T^*
  \ee
\be
  \sup\limits_{0\le t\le T}\norm[H^2]{u}\le C,\quad 0\le T<T^*
\ee
\end{proposition}

\proof  Differentiating the momentum equations in $(1.1)$ with respect to time $t$ yields
\be
\rho u_{tt} + \rho u\cdot\nabla u_t - \mu\triangle u_t -
(\mu + \lambda)\nabla\rm{div}u_t + \nabla p_t =
-\rho_t(u_t + u\cdot\nabla u) - \rho u_t\cdot\nabla u
\ee
 Taking the inner product of the above equation with $u_t$ in $L^2(\O)$ and integrating by parts, one gets
\be
\ba
& \frac{d}{dt}\int_{\O}\frac{1}{2}\rho u_t^2dx + \int_{\O}
(\mu|\nabla u_t|^2 + (\lambda + \mu)(\rm{div}u_t)^2)dx - \int_{\O}P_t\rm{div}u_tdx \\
& = -\int_{\O}(\rho u\cdot\nabla[(u_t + u\cdot\nabla u)u_t] +
\rho(u_t\cdot\nabla u)\cdot u_t)dx \ea
\ee

Due to $(2.18)$, the last term on the left-hand side of $(2.42)$ can be rewritten as
\be
\ba -\int_{\O}P_t\rm{div}u_tdx & =
\frac{d}{dt}\int_{\O}\frac{\g}{2}P(\rm{div}u)^2dx +
\int_{\O}\nabla P\cdot(u\rm{div}u_t)dx \\
& + \frac{\g}{2}\int_{\O}(-Pu\cdot\nabla(\rm{div}u)^2 +
(\g-1)P(\rm{div}u)^3)dx \ea
\ee

It follows from $(2.41)$ and $(2.42)$ that
\be
\ba & \frac{d}{dt}\int_{\O}(\frac{1}{2}\rho u_t^2 +
\frac{\g}{2}P(\rm{div}u)^2)dx +
\int_{\O}\mu|\nabla u_t|^2dx\\
& \le \int_{\O}(2\rho|u||u_t||\nabla u_t| + \rho|u||u_t||\nabla u|^2
+ \rho|u|^2|u_t||\nabla^2u|
+ \rho|u|^2||\nabla u||\nabla u_t| \\
& + \rho|u_t|^2|\nabla u| + |\nabla P||u||\nabla u_t| + \g P|u||\nabla u||\nabla^2u| + \g^2P|\nabla u|^3)dx\\
& \equiv \sum_{i=0}^{8}F_i \ea
\ee

Now, we estimate each $F_i$ separately, where the Sobolev inequality and H\"{o}lder inequality will
be used frequently.

\be
\ba
|F_1| & = \int_{\O}2\rho|u||u_t||\nabla u_t|dx\\
& \le C\norm[L^6]{u}\norm[L^3]{\rho^{1/2}u_t}\norm[L^2]{\nabla u_t}\\
& \le C\norm[L^2]{\rho^{1/2}u_t}^{\frac{1}{2}}\norm[L^2]{\nabla u_t}^{\frac{3}{2}}\\
& \le \ep\norm[L^2]{\nabla u_t}^2 + C\norm[L^2]{\rho^{1/2}u_t}^2\quad ,
\ea
\ee

where one has used $(2.27)$ and the interpolation inequality.

Similarly, it follows from $Lemma 2.1$ and $Proposition 2.4$ that
\be
\ba
|F_2| & = \int_{\O}\rho|u||u_t||\nabla u|^2dx \\
& \le C\norm[L^6]{u}\norm[L^6]{u_t}\norm[L^3]{\nabla u}^2\\
& \le C\norm[L^2]{\nabla u_t}\norm[L^2]{\nabla u}\norm[L^6]{\nabla u}\\
& \le C\norm[L^2]{\nabla u_t}\norm[L^6]{\nabla u}\\
& \le \ep\norm[L^2]{\nabla u_t}^2 + C\norm[H^2]{u}^2\quad ,
\ea
\ee

\be
\ba
|F_3| & = \int_{\O}\rho|u|^2|u_t||\nabla^2 u|dx\\
& \le \norm[L^3]{u^2}\norm[L^6]{u_t}\norm[L^2]{\nabla^2 u}\\
& \le \ep\norm[L^2]{\nabla u_t}^2 + C\norm[H^2]{u}^2\quad ,
\ea
\ee

\be
\ba
|F_4| & = \int_{\O}\rho|u|^2|\nabla u||\nabla u_t|dx \\
& \le C\norm[L^2]{\nabla u_t}\norm[L^6]{\nabla u}\norm[L^3]{u^2}\\
& \le C\norm[L^6]{\nabla u}\norm[L^2]{\nabla u_t}\\
& \le \ep\norm[L^2]{\nabla u_t}^2 + C\norm[H^2]{u}^2\quad ,
\ea
\ee

\be
\ba
|F_5| & = \int_{\O}\rho|u_t|^2|\nabla u|dx \\
& \le C\norm[L^2]{\rho u_t^2}\norm[L^2]{\nabla u}\\
& \le C\norm[L^4]{\rho^{1/2}u_t}^2\\
& \le \ep\norm[L^6]{u_t}^2 + C\norm[L^2]{\rho^{1/2}u_t^2}\quad ,
\ea
\ee

\be
\ba
|F_6| & = \int_{\O}|\nabla P||u||\nabla u_t|dx\\
& \le C\norm[L^2]{\nabla P}\norm[L^{\infty}]{u}\norm[L^2]{\nabla u_t}\\
& \le C\norm[H^2]{u}\norm[L^2]{\nabla u_t}\\
& \le \ep\norm[L^2]{\nabla u_t}^2 + C\norm[H^2]{u}^2\quad ,
\ea
\ee

\be
\ba
|F_7| & = \int_{\O}\g P|u||\nabla u||\nabla^2 u|dx \\
& \le C\norm[L^2]{\nabla^2 u}\norm[L^2]{\nabla u}\norm[L^{\infty}]{u}\\
& \le C\norm[L^2]{\nabla^2 u}\norm[L^{\infty}]{u}\\
& \le C\norm[H^2]{u}^2\quad ,
\ea
\ee

and finally,
\be
\ba
|F_8| & = \int_{\O}\g^2P|\nabla u|^3dx\\
& \le C\int_{\O}|\nabla u|^3dx \\
& \le C\norm[L^{\infty}(\O)]{\nabla u}\int_{\O}|\nabla u|^2dx\\
& \le C\norm[L^{\infty}(\O)]{\nabla u}\quad . \ea
\ee

Collecting all the estimates for $F_i$, we conclude
\be
\ba & \frac{d}{dt}\int_{\O}(\frac{1}{2}\rho u_t^2 +
\frac{\g}{2}P(\rm{div}u)^2)dx +
\int_{\O}\mu|\nabla u_t|^2dx\\
& \le 6\ep\int_{\O}|\nabla u_t|^2dx + C(\norm[L^2]{\rho^{1/2}u_t}^2
+ \norm[H^2]{u}^2 + \norm[L^2]{\nabla\rho}^2 +
\norm[L^{\infty}]{\nabla u}) \ea
\ee

Thanks to the compatibility condition:
\be
\rho_0(x)^{\frac{1}{2}}(\rho_0(x)^{\frac{1}{2}}u_t(t=0,x)
+ \rho_0^{\frac{1}{2}}u_0\cdot\nabla u_0(x) - \rho_0^{\frac{1}{2}}g) = 0
\ee
it holds that
\be
\rho_0(x)^{\frac{1}{2}}u_t(t=0,x) =
 \rho_0^{\frac{1}{2}}u_0\cdot\nabla u_0(x) - \rho_0^{\frac{1}{2}}g \in L^2(\O)
\ee

Therefore, for arbitrary small $\ep$, $(2.53)$ yields
\be
\sup\limits_{0\le t\le T}\norm[L^2]{\rho^{1/2}u_t(t)}^2 + \int_{Q_T}|\nabla u_t|^2dxdt \le C,
  \quad 0\le T<T^*
\ee
Moreover,
$$
 Lu = \rho u_t + \rho u\cdot\nabla u + \nabla P\
$$
thus,
$$\norm[H^2]{u}\le C(\norm[L^2]{\rho^{\frac{1}{2}}u_t} + \norm[L^6]{u}\norm[L^3]{\nabla u} + \norm[L^2]{\nabla P})
\le C(\norm[L^2]{\rho^{\frac{1}{2}}u_t} + \norm[L^2]{\nabla u}^{\frac{3}{2}}
\norm[H^2]{u}^{\frac{1}{2}} + \norm[L^2]{\nabla P})$$
Hence,
\be
\sup\limits_{0\le T< T^*}\norm[H^2]{u}^2\le C
\ee
Thus, $Proposition 2.5$ follows immediately.

\endproof

Finally, the following lemma gives bounds of the first order derivatives of the
density and the second derivatives of the velocity.

\begin{lemma}
Under the condition $(2.1)$, it holds that
$$\sup\limits_{0\le t\le T}(\norm[L^6]{\rho_t(t)} + \norm[W^{1,6}]{\rho})\le C,\quad 0\le T<T^*$$
$$\int_0^T\norm[W^{2,6}]{u(t)}^2dt\le C,\quad 0\le T<T_*$$
\end{lemma}

\proof It follows from $(2.55)$ and $(2.56)$ that
$$u_t\in L^2(0,T;L^6(\O)), \nabla u\in L^6(Q_T)$$
$$F\in L^2(0,T;L^6(\O))$$

Differentiating the mass equation in $(1.1)$ with
respect to $x_i$, and multiplying the resulting identity by $6|\p_i\rho|^4\p_i\rho$,
one gets after integration that
\be
\ba \p_t\int_{\O}|\p_i\rho|^6dx & =
-5\int_{\O}|\p_i\rho|^6\rm{div}udx -
6\int_{\O}\rho|\p_i\rho|^4\p_i\rho\p_i\rm{div}udx \\
& - 6\int_{\O}|\p_i\rho|^4\p_i\rho\p_iu\cdot\nabla\rho dx\\
& = -(B_1 + B_2 + B_3) \ea
\ee
Using $Lu = F + \nabla P$, one can estimate each term on the righthand side of $(2.57)$ as follows:

\be
|B_1(t)|\le 5\norm[L^{\infty}]{\nabla
u}(t)\int_{\O}|\p_i\rho|^6dx \le C\norm[L^{\infty}]{\nabla
u}(t)\int_{\O}|\nabla\rho|^6dx\quad ,
\ee

\be
|B_2(t)|\le
C\norm[L^{\frac{6}{5}}]{|\nabla\rho|^5}
(\norm[L^6]{\nabla P} + \norm[L^6]{F})\quad ,
\ee

\be
|B_4(t)|\le C\norm[L^{\infty}]{\nabla u}(t)\int_{\O}|\nabla\rho|^6dx\quad.
\ee

It follows from $(2.57)-(2.60)$ that
\be
\p_t\norm[L^6]{\nabla\rho} \le C(\norm[L^{\infty}]{\nabla u}(t) +
1)\norm[L^6]{\nabla\rho} + C\norm[L^6]{F}
\ee

Hence,
$$\sup\limits_{0\le t\le T}\norm[L^6]{\nabla\rho}\le C\quad.$$
Therefore, due to this, $(2.57)$ and interpolation inequality, one has
\be
\rho_t = -(u\cdot\nabla\rho + \rho\rm{div}u)\in L^{\infty}L^6\quad.
\ee
Finally, taking into account that
$$
Lu = F +\nabla P\in L^2L^6\quad,
$$
one has
\be
\int_0^T\norm[W^{2,6}(\O)]{u}^2dt \le C\quad.
\ee
This finishes the proof of $Lemma 2.6$.
\endproof

\section{Improved regularity of the density and the velocity}
In this section, we obtain some higher order regularity of the
density and the velocity. However, we may not deduce the $L^{\infty}H^1$ estimate
of $\nabla\rho$ directly just similar to $Lemma 2.4$ or $Lemma 2.6$, as
the $L^2$ norm of $\nabla^2 P$ can't be controlled by that of $\nabla^2\rho$ due to
the presence of vacuum, unless $\gamma$ is large is enough. In order to circumvent such
difficulties, we first need to improve the regularity of the pressure by observing that
$P$ satisfies a linear transport equation.

In fact, we have the following lemma.
\begin{lemma}
  \be
  \ba
& \norm[L^{\infty}H^2]{P} + \norm[L^{\infty}H^1]{P_t} + \norm[L^2L^2]{P_{tt}} \le C, \quad 0\le T<T^* \\
& \norm[L^{\infty}H^2]{\rho} + \norm[L^{\infty}H^1]{\rho_t} + \norm[L^2L^2]{\rho_{tt}} \le C,
\quad 0\le T<T^*
\ea
  \ee
\end{lemma}

\proof
For the proof of $(3.1)$, we will make use of the transport equation $(2.18)$ for the pressure
and the elliptic regularity of the system $Lu = F + \nabla P$ for the velocity $u$.

Indeed, it follows from the elliptic regularity that
\be
\norm[H^3]{u}\le C(\norm[H^1]{F} + \norm[H^1]{\nabla P})
\le C(\norm[H^1]{F} + \norm[L^2]{\nabla^2 P} + C)
\ee

Apply $D_{ij}$ to both side of $(2.18)$ to yield
\be
\ba
& (D_{ij}P)_t + D_{ij}u\cdot\nabla P + u\cdot\nabla D_{ij}P  +
D_iu\cdot\nabla D_jP + D_ju\cdot\nabla D_iP \\
& + \g D_{ij}P\rm{div}u + \g PD_{ij}\rm{div}u + \g(D_iPD_j\rm{div}u + D_jPD_i\rm{div}u)= 0
\ea
\ee

Multiplying $(3.3)$ by $2D_{ij}P$, one gets
\be
\ba
& \p_t(D_{ij}P)^2 + \rm{div}(|D_{ij}P|^2u) + (2\g-1)|D_{ij}P|^2\rm{div}u
+2D_{ij}PD_iu\cdot\nabla D_jP + 2D_{ij}PD_ju\cdot\nabla D_iP\\
& + 2\g PD_{ij}PD_{ij}\rm{div}u +
2\g D_iPD_{ij}PD_j\rm{div}u + 2\g D_jPD_{ij}PD_i\rm{div}u + +2D_{ij}PD_{ij}u\cdot\nabla P = 0
\ea
\ee

Integrating $()3.4$ over $\O$, yields
\be
\ba
& \p_t\int_{\O}|D_{ij}P|^2dx = -(2\g-1)\int_{\O}|D_{ij}P|^2\rm{div}udx -
2\int_{\O}D_{ij}PD_iu\cdot\nabla D_jPdx \\
& - 2\int_{\O}D_{ij}PD_ju\cdot\nabla D_iPdx - 2\g\int_{\O}PD_{ij}PD_{ij}\rm{div}udx \\
& - 2\g\int_{\O}D_iPD_{ij}PD_j\rm{div}udx -2\g\int_{\O}D_jPD_{ij}PD_i\rm{div}udx -
2\int_{\O}D_{ij}PD_{ij}u\cdot\nabla Pdx\\
& = -\sum_{i=1}^7P_i
\ea
\ee

Each term of $P_i$ can be estimated as follows
\be
|P_1,P_2,P_3| \le C\norm[L^{\infty}]{\nabla u}\int_{\O}|\nabla^2P|^2dx\quad,
\ee

\be
\ba
|P_4| & \le C\norm[L^2]{\nabla^2P}\norm[L^2]{D_{ij}\rm{div}u} \\
& \le C\norm[L^2]{\nabla^2P}(\norm[H^1]{F} + \norm[L^2]{\nabla^2P} + C)\\
& \le C\norm[L^2]{\nabla^2P}^2 + C\norm[H^1]{F}^2 + C\quad,
\ea
\ee

\be
\ba
|P_5,P_6,P_7| & \le C\norm[L^3]{D_iP}\norm[L^2]{\nabla^2P}\norm[L^6]{\nabla^2u}\\
& \le C\norm[L^2]{\nabla^2P}\norm[L^6]{\nabla^2u}\\
& \le C\norm[L^2]{\nabla^2P}^2 + C\norm[L^6]{\nabla^2u}^2\quad.
\ea
\ee
where one has used $lemma 2.6$.
Collecting $(3.5)-(3.8)$ yields
\be
\ba
& \p_t\int_{\O}|D_{ij}P|^2dx \le C(\norm[L^{\infty}]{\nabla u} + 1)\int_{\O}|\nabla^2P|^2dx \\
& + C(\norm[H^1]{F}^2 + \norm[L^6]{\nabla^2u}^2 + 1)
\ea
\ee

Using Gronwall's inequality and $P_0\in H^3$, $F\in L^2H^1$, $u\in L^2W^{2,6}$, one has
\be
\norm[L^{\infty}H^2]{P}\le C
\ee

As a consequence of $(2.18)$, $(3.10)$, $Lemma 2.1$ and $Proposition 2.4,2.5$, one has
\be
\norm[L^{\infty}H^1]{P_t} \le C
\ee

In view of $(3.10)-(3.11)$, we may apply the same technique to the mass equation to derive
\be
\norm[L^{\infty}H^2]{\rho} + \norm[L^{\infty}H^1]{\rho_t} \le C
\ee
Note that
$$
\rho_{tt} + \rho_t\rm{div}u + \rho\rm{div}u_t + u_t\cdot\nabla\rho + u\cdot\nabla\rho_t = 0
$$
$$
P_{tt} + \gamma P_t\rm{div}u + \gamma P\rm{div}u_t + u_t\cdot\nabla P + u\cdot\nabla P_t = 0
$$
then one has $\rho_{tt}\in L^2L^2$ and $P_{tt}\in L^2L^2$. Thus the lemma is proved
due to $(3.10)-(3.12)$, $Lemma 2.4$ and $Proposition 2.5$.
\endproof

In order to obtain high regularity of $(\rho,u)$, we need the following
improved estimate.
\begin{lemma}
\be
  \int_{Q_T}\rho u_{tt}^2dxdt + \sup\limits_{0\le t\le T}\int_{\O}|\nabla u_t|^2dx \le C,
  \quad 0\le T<T^*
\ee
\end{lemma}

\proof
Multiplying $(2.41)$ by $u_{tt}$, and integrating by parts, one gets that
\be \ba & \int_{\O}\rho u_{tt}^2dx + \int_{\O}\rho u\cdot\nabla
u_t\cdot u_{tt}dx +
\frac{d}{dt}\int_{\O}\frac{\mu}{2}|\nabla u_t|^2 + \frac{\lambda + \mu}{2}(\rm{div}u_t)^2dx \\
& = \int_{\O}P_t\rm{div}u_{tt}dx - \int_{\O}\rho_t(u_t +
u\cdot\nabla u)u_{tt}dx
- \int_{\O}\rho u_t\cdot\nabla u\cdot u_{tt}dx\\
\ea \ee

Note that
\be |\int_{\O}\rho u\cdot\nabla u_t\cdot u_{tt}dx| \le
\ep\int_{\O}\rho u_{tt}^2dx + C\int_{\O}\rho(u\cdot\nabla u_t)^2dx\quad, \ee

\be \ba |\int_{\O}\rho u_t\cdot\nabla u\cdot u_{tt}dx| & \le
\ep\norm[L^2]{\rho^{\frac{1}{2}}u_{tt}}^2 + C\norm[L^3]{\rho^{\frac{1}{2}}u_t}^2\norm[L^6]{\nabla u}^2\\
& \le \ep\norm[L^2]{\rho^{\frac{1}{2}}u_{tt}}^2 + C\norm[L^2]{\rho^{\frac{1}{2}}u_t}\norm[L^6]{u_t}\\
& \le \ep\norm[L^2]{\rho^{\frac{1}{2}}u_{tt}}^2 + C\norm[L^2]{\nabla u_t}\quad, \ea \ee

The first term of the right hand side of $(3.14)$ becomes
\be
\int_{\O}P_t\rm{div}u_{tt}dx  = \frac{d}{dt}\int_{\O}P_t\rm{div}u_tdx - \int_{\O}P_{tt}\rm{div}u_tdx
\ee
which can be estimated by
\be
\ba
& |\int_{\O}P_t\rm{div}u_t(T)dx| \le \frac{\mu}{8}\int_{\O}|\nabla
u_t|^2(T)dx + C\norm[L^2]{P_t}^2(T)\le \frac{\mu}{8}\int_{\O}|\nabla
u_t|^2(T)dx + C\\
& |\int_{\O}P_{tt}\rm{div}u_tdx|\le \norm[L^2]{P_{tt}}\norm[L^2]{\rm{div}u_t}
\le \norm[L^2]{P_{tt}}^2 + \norm[L^2]{\nabla u_t}^2
\ea
\ee

The second term of the righthand side of $(3.14)$, can be rewritten as

\be
\ba
& \int_{\O}\rho_t(u_t + u\cdot\nabla u)u_{tt}dx \\
& = \frac{d}{dt}\int_{\O}\rho_t(\frac{1}{2}|u_t|^2)dx - \int_{\O}\rho_{tt}(\frac{1}{2}|u_t|^2)dx
 + \int_{\O}\rho_t(u\cdot\nabla u)u_{tt}dx
\ea
\ee

Each term of the right hand side of $(3.19)$ can be estimated as follows
\be
\ba
|\int_{\O}\rho_t(\frac{1}{2}|u_t|^2)(T)dx| & = |\int_{\O}\rm{div}(\rho u)(\frac{1}{2}|u_t|^2)(T)dx| \\
& = |\int_{\O}\rho u\cdot\nabla (\frac{1}{2}|u_t|^2)(T)dx| \\
& \le \frac{\mu}{8}\int_{\O}|\nabla u_t|^2(T)dx + C(\mu)\norm[L^2]{\rho^{\frac{1}{2}}|u_t|(T)}^2\\
& \le \frac{\mu}{8}\int_{\O}|\nabla u_t|^2(T)dx + C
\ea
\ee

It follows from $Lu_t = F_t + \nabla P_t$ and the standard elliptic regularity theory that
\be
\norm[H^2]{u_t}\le C\norm[L^2]{F_t} + C\norm[L^2]{\nabla P_t}
\ee
A simple calculation based on the previous estimates shows that
\be
\ba
\norm[L^2]{F_t} & \le \norm[L^2]{\rho_tu_t} + \norm[L^2]{\rho u_{tt}} +
 \norm[L^2]{\rho_tu\cdot\nabla u} + \norm[L^2]{\rho u_t\cdot\nabla u} + \norm[L^2]{\rho u\cdot\nabla u_t}\\
& \le C(\norm[L^3]{\rho_t}\norm[L^6]{u_t} + \norm[L^2]{\rho u_{tt}} + \norm[L^3]{\rho_t}\norm[L^6]{\nabla u}
+ \norm[L^6]{u_t}\norm[L^3]{\nabla u} + \norm[L^2]{\nabla u_t})\\
 & \le C(\norm[L^2]{\rho^{\frac{1}{2}} u_{tt}} + \norm[L^2]{\nabla u_t} + 1)
\ea
\ee

Accordingly, the second term of righthand side of $(3.19)$ becomes
\be
\ba
|\int_{\O}\rho_{tt}(\frac{1}{2}|u_t|^2)dx| &  = |\int_{\O}\rm{div}(\rho_tu + \rho u_t)
(\frac{1}{2}|u_t|^2)dx|\\
& = |\int_{\O}(\rho_tu + \rho u_t)\nabla(\frac{1}{2}|u_t|^2)dx|\\
& \le C\norm[L^3]{\rho_t}\norm[L^{\infty}]{u}\norm[L^6]{u_t}\norm[L^2]{\nabla u_t}
 + \int_{\O}\rho |u_t|^2|\nabla u_t|dx\\
& \le C\norm[L^2]{\nabla u_t}^2 + C\norm[L^3]{\rho^{\frac{1}{2}}u_t}^2\norm[L^3]{\nabla u_t}\\
& \le C\norm[L^2]{\nabla u_t}^2 + C\norm[L^2]{\rho^{\frac{1}{2}}u_t}\norm[L^6]{\rho^{\frac{1}{2}}u_t}
\norm[L^3]{\nabla u_t}\\
& \le C\norm[L^2]{\nabla u_t}^2 + C\norm[L^2]{\nabla u_t}\norm[L^3]{\nabla u_t}\\
& \le C\norm[L^2]{\nabla u_t}^2 + C\norm[L^2]{\nabla u_t}\norm[H^2]{u_t}\\
& \le C\norm[L^2]{\nabla u_t}^2 + C\norm[L^2]{\nabla u_t}
(\norm[L^2]{\rho^{\frac{1}{2}} u_{tt}} + \norm[L^2]{\nabla u_t} + 1)\\
& \le C\norm[L^2]{\nabla u_t}^2 + \ep\norm[L^2]{\rho^{\frac{1}{2}}u_{tt}}^2 + C
\ea
\ee
where we use $(3.23)$ and $(3.24)$.
We write the last term of righthand side of $(3.21)$ as
\be
\ba
& \int_{\O}\rho_t(u\cdot\nabla u)u_{tt}dx \\
& = \frac{d}{dt}\int_{\O}\rho_t(u\cdot\nabla u)u_tdx - \int_{\O}\rho_{tt}(u\cdot\nabla u)u_tdx
-\int_{\O}\rho_t(u_t\cdot\nabla u)u_tdx - \int_{\O}\rho_t(u\cdot\nabla u_t)u_tdx
\ea
\ee

Observe that

\be
\ba
|\int_{\O}\rho_t(u\cdot\nabla u)u_tdx| & \le \norm[L^3]{\rho_t}\norm[L^2]{u\cdot\nabla u}
\norm[L^6]{u_t} \\
& \le \frac{\mu}{8} \norm[L^2]{\nabla u_t}^2 + C\quad,
\ea
\ee

and
\be
\ba
|\int_{\O}\rho_{tt}(u\cdot\nabla u)u_tdx|  & \le \norm[L^2]{\rho_{tt}}
\norm[L^3]{u\cdot\nabla u}\norm[L^6]{u_t} \\
& \le C\norm[L^2]{\rho_{tt}}\norm[L^6]{u_t}\\
& \le C\norm[L^2]{\rho_{tt}}^2 + C\norm[L^2]{\nabla u_t}^2\quad,
\ea
\ee

\be
\ba
|\int_{\O}\rho_t(u_t\cdot\nabla u)u_tdx| & \le \norm[L^2]{\rho_t}\norm[L^3]{|u_t|^2}
\norm[L^6]{\nabla u}\\
& \le C\norm[L^2]{\nabla u_t}^2\quad,
\ea
\ee

and
\be
\ba
|\int_{\O}\rho_t(u\cdot\nabla u_t)u_tdx| & \le
\norm[L^3]{\rho_t}\norm[L^{\infty}]{u}\norm[L^2]{\nabla u_t}\norm[L^6]{u_t}\\
& \le C\norm[L^2]{\nabla u_t}\norm[L^6]{u_t}\\
& \le C\norm[L^2]{\nabla u_t}^2
\ea
\ee

It follows from $Lemma 3.1$ and $Proposition 2.5$ that
$$(\rho_t,P_t)\in L^{\infty}H^1, (\rho_{tt},P_{tt})\in L^2L^2, \nabla u_t\in L^2L^2$$

In view of regularity $(1.15)$, there exist a sequence $\ep_i$, such that $\ep_i\rightarrow 0, \ep_i>0$, and
\be
\norm[H^{1}]{u_t(\ep_i)}\le \norm[L^{\infty}(0,T_*;H_0^1(\O))]{u_t}\le C(\norm[H^3]{u_0},\norm[H^3]{\rho_0}
,\norm[H^3]{P_0})
\ee
Collecting all the estimates $(3.14)-(3.29)$, integrating over $(\ep_i,T)$, accordingly
\be
\ba
& \int_{\ep_i}^T\int_{\O}\rho u_{tt}^2dxdt + \int_{\O}\frac{\mu}{8}|\nabla u_t(T)|^2 +
\frac{\lambda + \mu}{2}(\rm{div}u_t(T))^2dx \\
& \le 3\ep\int_{\ep_i}^T\int_{\O}\rho u_{tt}^2dxdt + C\int_{\ep_i}^T
(\norm[L^2]{P_{tt}}^2 + \norm[L^2]{\rho_{tt}}^2
 + \norm[L^2]{\nabla u_t}^2 + \norm[L^2]{\rho^{\frac{1}{2}}u_t}^2 + 1)dt\\
& + \int_{\O}\frac{\mu}{8}|\nabla u_t(\ep_i)|^2 +
\frac{\lambda + \mu}{2}(\rm{div}u_t(\ep_i))^2dx\\
& \le 3\ep\int_{0}^T\int_{\O}\rho u_{tt}^2dxdt + C\int_{0}^T
(\norm[L^2]{P_{tt}}^2 + \norm[L^2]{\rho_{tt}}^2
 + \norm[L^2]{\nabla u_t}^2 + \norm[L^2]{\rho^{\frac{1}{2}}u_t}^2 + 1)dt\\
& + C(\norm[H^3]{u_0},\norm[H^3]{\rho_0},\norm[H^3]{P_0})
\ea
\ee
The righthand of $(3.30)$ is independent of $\ep_i$. Therefore, letting $\ep_i$ go to $0$
and choosing $\ep$ small enough, we complete the proof of lemma $3.2$.

\endproof

Finally, we have
\begin{lemma}
  \be
  \norm[L^{\infty}H^3]{\rho} + \norm[L^{\infty}H^3]{P} + \norm[L^{\infty}H^3]{u}\le C
  \ee
\end{lemma}

\proof
It follows from $(3.1)$ and $(3.15)$ that
\be
F = \rho u_t + \rho u\cdot\nabla u\in L^{\infty}H^1, \quad \nabla P\in L^{\infty}H^1
\ee
which gives
\be
Lu = F + \nabla P\in L^{\infty}H^1
\ee
As a consequence,
\be
\norm[L^{\infty}H^3]{u}\le C
\ee
Therefore,
\be
Lu_t = F_t + \nabla P_t\in L^2L^2
\ee
which implies
\be
u_t\in L^2H^2, F\in L^2H^2
\ee

By an estimate similar to lemma $3.1$, one can derive the high regularity of pressure $P$,
it holds that
\be
\norm[L^{\infty}H^3]{P}\le C
\ee

In view of the mass equation, one can show that
\be
\norm[L^{\infty}H^3]{\rho}\le C
\ee
\endproof

This will be enough to extend the classical solutions of $(\rho,u)$ beyond $t\ge T^*$.

In fact, in view of Lemma $3.1 - 3.3$, the functions
$(\rho,P,u)|_{t = T^*} = \lim_{t\rightarrow T^*}(\rho,P,u)$ satisfy the conditions
imposed on the initial data $(1.13)-(1.14)$ at the time $t=T^*$ Furthermore,
$$
\rho u_t + \rho u\cdot\nabla u \in L^{\infty}H_0^1
$$
\be
-Lu + \nabla P|_{t = T^*} = \lim_{t\rightarrow T^*}
(\rho u_t + \rho u\cdot\nabla u)
\triangleq \rho g|_{t = T^*}\quad,
\ee
where $g|_{t = T^*}\in H_0^1(\O)$ and $\rho^{\frac{1}{2}}g|_{T^*}\in L^2$. Therefore,
we can take $(\rho,P,u)|_{t = T^*}$ as the initial data and apply the local existence theorem
\cite{K3} to extend our local classical solution beyond $T^*$. This contradicts the assumption on $T^*$.

Note that a few modifications can be applied for both
periodic case and $\O=R^3$, so theorem $1.2$ holds.

{\bf Acknowledgement}  This research is supported in part by Zheng Ge Ru Foundation,
Hong Kong RGC Earmarked Research Grants CUHK4028/04P, CUHK4040/06P, CUHK4042/08P, and
the RGC Central Allocation Grant CA05/06.SC01.

\end{document}